\documentclass[floatfix,twocolumn,aps,pra,superscriptaddress,showpacs]{revtex4-1}
\usepackage[utf8x]{inputenc}
\usepackage{default}
\usepackage{amssymb}
\usepackage{color}
\usepackage{amsmath}
\usepackage{mathrsfs}
\usepackage{graphicx}
\usepackage{subfig}
\usepackage{epsfig}

\begin{document}
\title{Interaction of solitons and the formation of bound states in the generalized Lugiato-Lefever equation}
\author{P. Parra-Rivas$^{1,2}$, D. Gomila$^{2}$, P. Colet$^{2}$,  L. Gelens$^{1,3}$}

\affiliation{ $^1$Applied Physics Research Group, APHY, Vrije Universiteit Brussel, 1050 Brussels Belgium.\\$^{2}$Instituto de F\'{\i}sica Interdisciplinar y
  Sistemas Complejos, IFISC (CSIC-UIB),Campus Universitat de les Illes
  Balears, E-07122 Palma de Mallorca, Spain.\\$^3$ Laboratory of Dynamics in Biological Systems, KU Leuven Department of Cellular and Molecular Medicine, University of Leuven, B-3000 Leuven, Belgium.}


\begin{abstract}
 Bound states, also called soliton molecules, can form as a result of the interaction between individual solitons. This interaction is mediated through the tails of each soliton that overlap with one another. When such soliton tails have spatial oscillations, locking or pinning between two solitons can occur at fixed distances related with the wavelength of these oscillations, thus forming a bound state. In this work, we study the formation and stability of various types of bound states in the Lugiato-Lefever equation by computing their interaction potential and by analyzing the properties of the oscillatory tails. Moreover, we study the effect of higher order dispersion and noise in the pump intensity on the dynamics of bound states. In doing so, we reveal that perturbations to the Lugiato-Lefever equation that maintain reversibility, such as fourth order dispersion, lead to bound states that tend to separate from one another in time when noise is added. This separation force is determined by the 
shape of the envelope of the interaction potential, as well as an additional 
Brownian ratchet effect. In systems with broken reversibility, such as third order dispersion,
this ratchet effect continues to push solitons within a bound state apart. However, the force 
generated by the envelope of the potential is now such that it pushes the solitons towards each other, 
leading to a null net drift of the solitons. 
\end{abstract}

\maketitle

\section{Introduction}\label{sec:inter:intro}

Localized dissipative structures (LSs) are a particular type of emergent structures appearing in spatially extended systems out of the thermodynamical equilibrium,
i.e.  systems that are internally dissipative and externally driven, commonly known as dissipative systems. 
They can be understood as part of one state embedded in a background of a different state and are formed due to a double balance between
nonlinearity and dispersion (or diffraction) on the one hand, and 
driving and dissipation on the other \cite{Akhmediev_GTS}. When they consist of a single localized element, such as a single peak or a localized patterned patch, LSs are also called dissipative solitons or simply solitons.
They are unique once the system parameters are fixed and they can exhibit multistability with other LSs \cite{Akhmediev_GTS}. These structures can display a variety 
of dynamical regimes such as periodic oscillations \cite{Umbanhowar,Firth,Vanag_Epstein,leo_lendert}, chaos \cite{leo_lendert,Michaelis},
or excitability \cite{gomila}, and they are found in a wide variety of systems in nature, such as chemical reactions, plant ecology, biology and nonlinear optics \cite{Cross_Hohenberg,Murray,Hoyle}. Moreover, several single LSs can also coexist and form bound states (BSs) that keep a stable separation distance between them \cite{Malomed91,Malomed93,Malomed94,Akhmediev,Akhmediev_Soto_PRL,Barashenkov,Schapers}.

In nonlinear optics, LSs arise in driven nonlinear optical cavities such as Fabry-Perot interferometers \cite{lugiato_spatial_1987}, fiber ring cavities \cite{Haelterman,Leo_NaturePhot}, and microresonators \cite{Coen_comb,Chembo_comb}. In this context, LSs are well described by a mean-field model, which in the case of Kerr nonlinearities reduces to the Lugiato-Lefever (LL) equation \cite{lugiato_spatial_1987}. In the last decade, this model has sparked new interest as it has been found to describe accurately the formation and dynamics of Kerr frequency combs in microresonators \cite{Coen_comb,Chembo_comb}. Optical frequency combs in microresonators show great promise thanks to their wide range of applications and their potential to be integrated on-chip \cite{kippen,delhaye,Okawachi,hansch,Scott,Herr_kippenberg}. Therefore, understanding the formation of single solitons, patterns, and bound states, is a crucial step in gaining new insights into optical frequency combs.

Several works by Malomed \cite{Malomed91,Malomed93,Malomed94}, Akhmediev \cite{Akhmediev,Akhmediev_Soto_PRL}, and Barashenkov \cite{Barashenkov} have 
theoretically studied BSs in perturbed versions of the Nonlinear Schr\"odinger (NLS) equation using a variety of techniques,
such as calculating effective interaction potentials. Even in non-integrable equations, which do not allow finding a closed analytical 
expression for the interaction potential, one can use numerical solutions of solitons to calculate such a potential. In \cite{Malomed91},
it was found that equilibrium distances of BSs correspond to the maxima and minima of that potential. The maxima correspond with the unstable 
equilibrium separations, while the minima indicate the stable ones. The interaction in the presence of so-called "skew" terms (terms breaking spatial reversibility) 
has also been studied \cite{Malomed93,Malomed94}. In this particular case, the maxima of the potential, instead of the minima, determine the stable equilibrium separation distances for BSs \cite{Malomed93}. In even earlier works, it was shown that locking of solitons can be produced via
radiative interaction or dispersive wave emission \cite{Mollenauer,Kaup}. Interaction of solitons in dissipative systems has also
been studied using perturbation theory in the modified Swift-Hohenberg equation \cite{Tlidi_SHE,Tlidi_SHE_interaction}. We also note that BSs have been 
studied in two and three spatial dimensions \cite{Malomed98},  and that the solitons within a BS can undergo temporal oscillatory instabilities and move with respect to one another \cite{soto_crespo}.
Experimentally, BSs have been observed in spatial optical cavities \cite{Schapers}, and more recently 
in microresonators \cite{Herr_kippenberg,Wang_leo_jae_coen,Wang_leo_jae_coen_Optica,Brasch,Vahala,DelHaye}.

In this paper, using similar techniques as in \cite{Malomed91,Malomed93,Malomed94,Akhmediev,Barashenkov}, we study how different LSs can interact and form 
BSs in the generalized LL equation due to their interaction via the oscillatory tails in the soliton profiles. To do so, we focus 
on the anomalous group velocity dispersion (GVD) regime. In Section~\ref{sec:interac:LS}, we introduce the standard LL model and review how 
different types of LSs form. Next, in Section~\ref{sec:BS}, we show a variety of BSs and discuss the general mechanism by which the solitons
can interlock at different separation distances. In the next Section,  we then calculate an interaction potential using variational principles, 
and analyze how the shape of the tails of the LSs changes. Using both approaches, we discuss for which control parameters BSs of various
separation distances can form. Afterwards, in Section~\ref{sec:interac:bif}, we present different types of bifurcation structures for BSs.
Section~\ref{sec:interac:HOT} is 
devoted to the generalization of these results to the LL equation in the presence of high order
dispersion effects, in particular third order dispersion (TOD) and fourth order dispersion (FOD), where we also 
perform simulations in the presence of noise analyzing how two solitons jump between neighboring stable distances (Section 7). 
Finally, in Section \ref{sec:interac:conclusion}, we end with some concluding remarks.

\section{Localized structures in the Lugiato-Lefever equation}\label{sec:interac:LS}

The standard LL equation with one spatial dimension in the anomalous GVD regime reads
\begin{equation}\label{LLE}
 \partial_tA=-(1+i\theta)A+i\partial_x^2A+iA|A|^2+\rho
\end{equation}
where, $\rho$, $\theta$  are control parameters representing the normalized
injection amplitude and the frequency detuning, respectively \cite{lugiato_spatial_1987,Haelterman}.

Steady state solutions of (\ref{LLE}) satisfy the following ordinary differential equation
\begin{equation}\label{LLEsteady}
i \displaystyle\frac{d^2A}{dx^2}(x)-(1+i\theta)A(x)+iA(x)|A(x)|^2+\rho=0, 
\end{equation}
and they can be homogeneous steady states (HSS), or non-uniform solutions (with a spatial dependence) consisting of periodic patterns or LSs \cite{Coullet}. 

The HSS solutions $A_0$ are given by the classic cubic equation of dispersive optical bistability, namely
\begin{equation}\label{HSS}
 I_0^3-2\theta I_0^2+(1+\theta^2)I_0=\rho^2
\end{equation} 
where $I_0\equiv|A_0|^2$. For $\theta<\sqrt{3}$, (\ref{HSS}) is monovaluate and hence the system is monostable. However, if 
$\theta>\sqrt{3}$, (\ref{HSS}) is trivaluate and $A_0$ has three branches of solutions forming a S-shape bifurcation diagram.
The different branches meet at saddle-nodes SN$_{hom,1}$ and SN$_{hom,2}$ located at
\begin{equation}
 I_{b,t}=\frac{2\theta}{3}\pm\frac{1}{3}\sqrt{\theta^2-3}.
\end{equation}

In terms of the real and imaginary parts of the field A, the HSSs $A_0$ is given implicitly by
\begin{equation}
\left[\begin{array}{c}
{\rm Re}[A_0] \\ {\rm Im} [A_0] \end{array}\right]=\left[\begin{array}{c}
\displaystyle\frac{\rho}{1+(I_0-\theta)^2} \\ \displaystyle\frac{(I_0-\theta)\rho}{1+(I_0-\theta)^2}\end{array}\right].
\end{equation}

\begin{figure*}
\centering
\includegraphics[scale = 1]{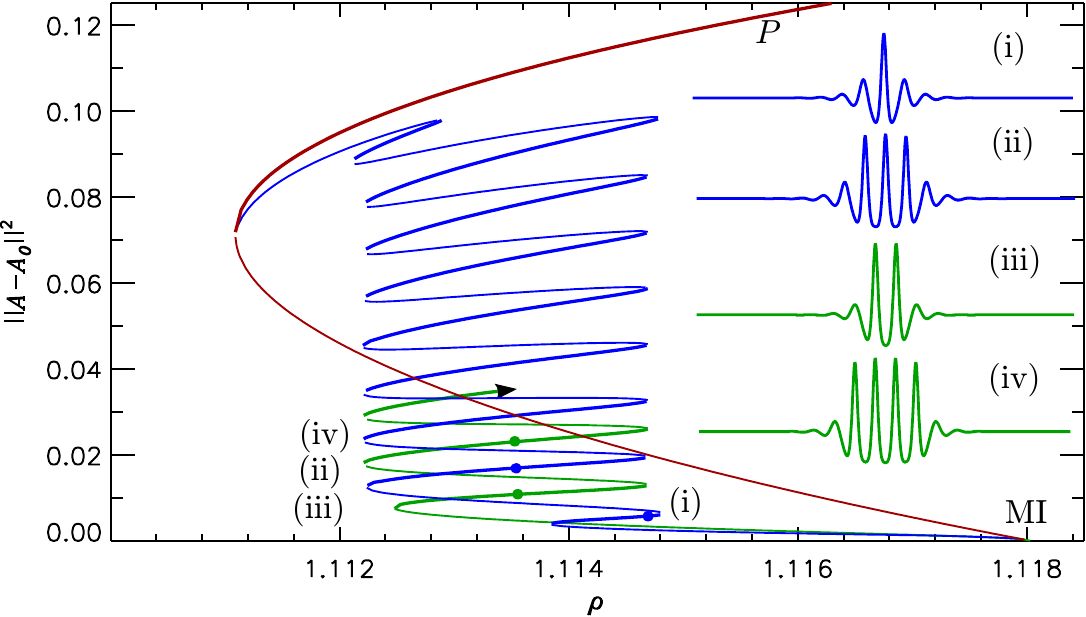}
\caption{Homoclinic snaking diagram and LSs for $\theta=1.5$ and $L=160$. In blue (green), the branches of solutions of LSs with odd (even) number of peaks; in red the branches corresponding to the pattern solution $P$. Thicker (thinner) branches correspond to stable (unstable) solutions. Examples of solution profiles are shown in panels (i) - (iv). 
} 
\label{Hsnake}
\end{figure*}

For $\theta<\sqrt{3}$, the homogeneous solution, $A_0$, becomes unstable to perturbations of the form $\propto{\rm e}^{\Omega t-ikx}$ at $I_0=I_c=1$, 
with critical wavenumber $k=k_c=\sqrt{2-\theta}$. Above such threshold the real part of $\Omega$ is positive and perturbations grow. At $I_c$ a pattern solution is created either supercritically ($\theta<41/30$) 
or subcritically ($\theta>41/30$) \cite{lugiato_spatial_1987}.  If the intracavity intensity $I_0$ is above the modulation instability (MI) threshold, $I_c=1$, the HSS is unstable to a range of wavevectors, with $k=k_u=\sqrt{2I_0-\theta}$ the one growing the fastest. For $\sqrt{3}<\theta<2$, while the homogeneous solution is trivalued, the lower branch ($A_0^b$) has the same instability threshold and critical wavenumber. 

In the regime where patterns arise subcritically, the bistability between the low intensity solution ($A_0^b$, zero in Fig.~\ref{Hsnake}) and the pattern solution 
($P$, red branch in Fig.~\ref{Hsnake}) allows the formation of LSs \cite{Woods_1999,gomila_Scroggi,Burke_2006}.
Together with the pattern, two branches of LSs bifurcate from the MI. One branch includes profiles with an odd number of peaks (blue curve), and the other
one contains profiles with an even number of peaks (green curve). These two branches of solutions are intertwined in a so-called {\it homoclinic snaking} bifurcation structure: 
a sequence of saddle-node bifurcations cause the branches to oscillate back and forth across a parameter range
called the snaking or pinning region \cite{Woods_1999,Burke_2006}. The LSs repeatedly add peaks on either side symmetrically at each back-and-forth oscillation, increasing the width of the LS by 
two wavelengths $2\pi/k_c$, until filling the whole size and connecting to the pattern solution $P$. This type of bifurcation structure in the LL equation is shown in Fig.~\ref{Hsnake} for a domain size $L=160$ and for a fixed value of the detuning $\theta=1.5$, and was discussed in more detail in \cite{Parra_Rivas_2}. 
 
\begin{figure}
\centering
\includegraphics[scale = 0.8]{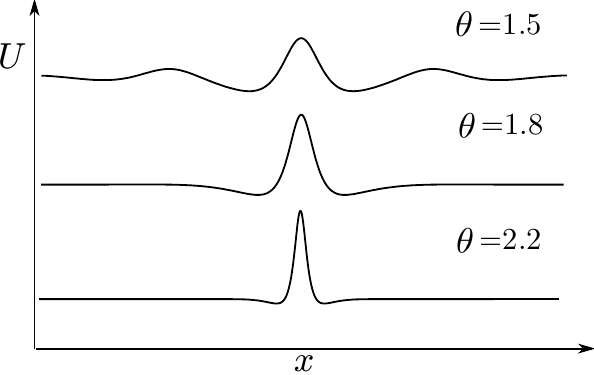}
\caption{Spatial profile of solitons at different values of the detuning: $(\theta, \rho) = (1.5,1.11445), (1.8,1.27), (2.2,1.5)$. }
\label{ProfilesDet}
\end{figure}

For higher values of the detuning ($\theta>2$), a region of operation that is more commonly used in frequency comb generation in microresonators, 
the situation is considerably different. The low intensity HSS, $A_0^b$, is now stable all the way up to SN$_{hom,1}$, thus
$I_{b}=\frac{2\theta}{3}+\frac{1}{3}\sqrt{\theta^2-3}$. When increasing the detuning the typical soliton becomes sharper and the oscillatory
tails decrease in amplitude (see Fig. \ref{ProfilesDet}). Now $I_c=1$ corresponds to a Belyakov-Devaney transition \cite{Parra_Rivas_2}. 
For $I_0 < I_c$, LSs still maintain oscillatory tails,
although they are highly damped. However, as soon as $I_0>I_c$, the tails become monotonic.

\section{Origin of bound states}\label{sec:BS}

\begin{figure}
\centering
\includegraphics[scale = 0.9]{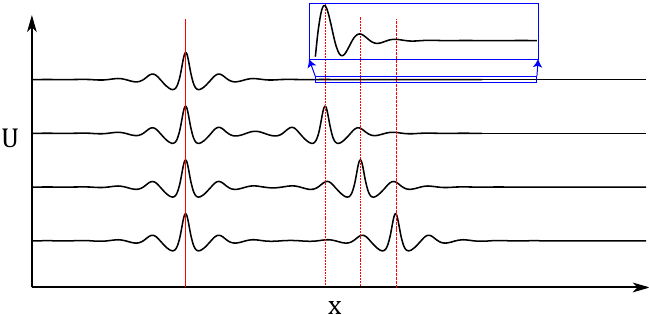}
\caption{Spatial profiles of a single soliton and several BSs consisting of two such solitons for $\theta=1.5$ and $\rho = 1.11445$. The red lines indicate the maxima of the solitons within the BSs and closely correlate with the maxima of the oscillatory tails of the single soliton where they interlock (see zoom of the tails in the inset).
} 
\label{BS_sketch}
\end{figure}

In Fig. \ref{BS_sketch}, we show a single stable soliton, as well as two different stable BSs consisting of two solitons bound at different separation distances. The maxima of all solitons are indicated by red lines, showing that the maxima of each soliton in the BS closely correlate with the maxima of the oscillatory tails of the single soliton. Indeed, it was analyzed in a general way that different LSs interact through their tails and can form bound states in this way \cite{Elphick,Aranson}. Moreover, it was shown that such tail interaction is not only important between different LSs, but LSs can also interact with system boundaries through their tails \cite{Kozyreff1,Kozyreff2}. 

This general mechanism of forming BSs through tail interaction has clear implications in the context of the LL equation. In Fig. \ref{ProfilesDet}, we illustrated that oscillatory tail amplitude damps increasing the frequency detuning $\theta$, weakening the interaction between solitons. As a result, BSs cannot be strongly locked together and are very sensitive to perturbations. The formation of BSs becomes even more cumbersome for 
values of the frequency detuning larger than 2 where tails become monotonic. The interaction via monotonic tails leads to a continuous, albeit very slow, movement of the solitons, so that they no longer form any stable BSs. In recent experiments, however, BSs have been found for $\theta > 2$, showing that the standard LL equation does not capture all of the dynamics. 

Any perturbation to the LL equation that introduces oscillatory tails into the soliton profile would be sufficient to explain the formation of BSs. In what follows, we will first discuss in detail BSs in the standard LL equation for $\theta < 2$. Next, we will extend our study to BSs in the generalized LL equation in the presence of higher order dispersion effects, which is one way to introduce oscillatory tails \cite{GelensPRL,Gelens_NL}.

\section{Interaction potential and oscillatory tails}\label{sec:interac:potential}

\subsection{Variational methods}
Here, using similar variational methods as in \cite{Malomed91,Malomed93,Malomed94,Akhmediev,Barashenkov}, we study how solitons interact and form BSs. The LL
equation  with one spatial dimension has the following generalized action functional:
\begin{equation}\label{S_chap}
 \mathcal{S}[A,\bar{A}]\equiv\int_{\mathbb{R}}e^{2t}L[A,\bar{A}]dt=\displaystyle\int_{\mathbb{R}}e^{2t}\displaystyle\int_{\mathbb{R}}\mathcal{L}[A,\bar{A}]dxdt,
\end{equation}
where $\bar{A}$ is the complex conjugate of the field $A$. The time-dependent exponential is linked to the dissipative nature of the system \cite{Barashenkov}.
In (\ref{S_chap}), the Lagrangian density is given by \cite{Firth_lagran}:
\begin{multline}\label{lagrangian_chap}
 \mathcal{L} [A,\bar{A}]\equiv\frac{i}{2}\left(\bar{A}\partial_t A-A\partial_t\bar{A}\right)\\-\partial_xA\partial_x\bar{A}-i
\rho\left(\bar{A}-A\right)-\theta A\bar{A}+\frac{1}{2}\bar{A}^2 A^2.
\end{multline}

In this framework the LL equation corresponds to the Euler-Lagrange equation, derived from the least action principle defined by: 
\begin{equation}\label{first_variation_chap}
\frac{\delta \mathcal{S}[A,\bar{A}]}{\delta \bar{A}} =0  
\end{equation}
where $\delta$ stands for functional derivative.

The Hamiltonian density is,
\begin{equation}\label{H_chap}
 \mathcal{H}[A,\bar{A}]=\partial_xA\partial_x\bar{A}+i\rho\left(\bar{A}-A\right)+\theta A\bar{A}-\frac{1}{2}\bar{A}^2 A^2,
\end{equation}
and the interaction Hamiltonian density is given only by the last term 
\begin{equation}\label{Hint_chap}
 \mathcal{H_{\rm I}}[A,\bar{A}]=-\displaystyle\frac{1}{2}A^2\bar{A}^2.
\end{equation}
The Hamiltonian function is then defined by 
 \begin{equation}
H[A,\bar A]=\displaystyle\int_{\mathbb{R}}\mathcal{H}[A,\bar A]dx.
\end{equation}
We now use this formalism to derive effective interaction potentials of two solitons that are separated by a distance $z$. This potential is determined 
by the overlapping integral between the tail of one soliton and the core of the other soliton. We consider that the BS formed by 
two solitons that are widely separated can be described by the ansatz
\begin{equation}\label{ansatz_interac}
A(x,z) = A_-(x)+A_+(x)-A_0,
\end{equation}
where  $A_-(x)=A_{sol}(x-z/2)$ represents a soliton displaced by a distance $z/2$ to the left of the center of the domain at $x=0$, 
and $A_+=A_{sol}(x+z/2)$ a soliton displaced by a distance $z/2$ to the right, with $z$ being a time-dependent free parameter \cite{Barashenkov}.
 In Ref.~\cite{Akhmediev_Soto_PRL} a different ansatz including the phase difference between solitons was considered. In the case considered here the phase 
 is fixed by the external pump and there is no need of such free parameter.
As there is not a known exact analytical solution for the soliton in the LL equation, we use the stationary solution profiles obtained numerically via a Newton method to calculate the potential. 

\subsection{Full Hamiltonian}
Using the ansatz (\ref{ansatz_interac}), the action functional depends on $z$, i.e. $\mathcal{S}(z) \equiv \mathcal{S}[A(z),\bar{A}(z)]$. This time independent ansatz implies that the kinetic terms in the Lagrangian density (\ref{lagrangian_chap}) vanish and thus the action corresponds to the integral of the Hamiltonian density (\ref{H_chap}) \cite{Barashenkov}. The extreme of the action correspond to stationary equilibrium distances or pinning distances:
\begin{equation}
 \delta_{z}\mathcal{S}(z)=
 \displaystyle\frac{d}{dz}\displaystyle\int_{\mathbb{R}} e^{2t}H(z)dt=0.
\end{equation}
which is equivalent to 
\begin{equation}\label{EL1}
 \displaystyle\frac{dH}{dz}=0.
\end{equation}
This equation has the form of a constraint and its solutions $z=z_n$ correspond to the equilibrium separation distances for the BSs. $H$ as a function of $z$ defines an effective potential 
\begin{equation}
\label{effec_pot}
 U_{H}(z)=H[A(z)]=\displaystyle\int_{\mathbb{R}}\mathcal{H}[A(z)]dx.
\end{equation}

\subsection{Interaction Hamiltonian}
Another constraint equation similar to (\ref{EL1}) can be obtained using the interaction Hamiltonian (\ref{Hint_chap})
\begin{equation}
 H_{I}=-\displaystyle\frac{1}{2}\int_{\mathbb{R}}A^2\bar{A}^2dx=-\displaystyle\frac{1}{2}\int_{\mathbb{R}}|A|^4dx,
\end{equation}
instead of the complete one given by (\ref{H_chap}). This Hamiltonian defines an effective interaction potential
\begin{equation}
\label{effec_pot_int}
 U_{H_{I}}(z)=H_{I}[A(z)]=-\displaystyle\frac{1}{2}\int_{\mathbb{R}}|A(x,z)|^4dx.
\end{equation}

\subsection{Effective potentials}
\begin{figure}[!t]
\centering
\includegraphics[scale = 0.8]{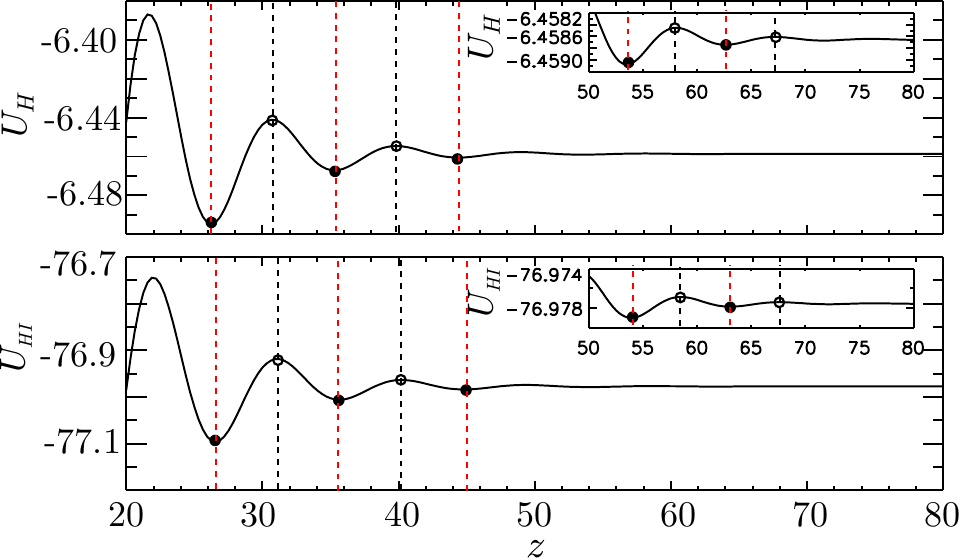}
\caption{Effective potential $U_H$ (a) using the full Hamiltonian and  $U_{H_I}$ (b) using the interaction Hamiltonian, both as a function of separation distance $z$ between soliton peaks. Parameters as in Fig.~\ref{BS_sketch}.}
\label{Poten}
\end{figure}

Fig. \ref{Poten} shows the potential corresponding to the soliton shown in Fig. \ref{BS_sketch}, calculated using the full Hamiltonian density (\ref{H_chap}) (top)
and the interaction Hamiltonian density (\ref{Hint_chap}) (bottom),  integrated over the range $[-L/2,L/2]$. In both cases the potentials oscillate in $z$ with a fixed 
period or wavelength $\Lambda_U$ and decays for increasing values of $z$. The insets in Fig.  \ref{Poten} show that even at large 
distances the oscillations are still present. The minima (maxima) of the potential correspond to stable (unstable)
equilibrium separations $z_n^s$ ($z_n^u$), and therefore to stable or unstable BSs. These minima (maxima) are indicated by $\bullet$
($\circ$) and by the red (black) dashed lines. The numerical values for some of these stable points using both calculated 
potentials are shown in Table~\ref{tab_inter1}. For comparison we have also added the exact separation distances
calculated using numerical time-evolution simulations. Although the agreement is not perfect, the differences are relatively small 
and of the order of those found in \cite{Barashenkov}, where a similar equation was used. They can be attributed to the approximation
(\ref{ansatz_interac}). The periodicity of this potential $\Lambda_U=z^s_{n+1}-z^s_n$, i.e. the difference between the position of two
consecutive local minima (or maxima) can be calculated for every set of control parameters. Here, we find that $\Lambda_{U}\in(9.1,9.2)$.

\begin{table}
\centering
\begin{tabular}{llll}
\hline\noalign{\smallskip}
Pinning dist. & $U_{H}$ &  $U_{H_{I}}$ & Exact  \\
\hline
$z^s_1$ & 35.2979  & 35.6575   &36.3181\\
\hline
$z^s_2$& 44.3926  & 44.7891 & 45.4856 \\
\hline
$z^s_3$ & 53.5116 & 53.931  & 54.6523 \\
\hline
$z^s_5$ & 80.0  & 80.0&    80.0 \\
\hline
$z^s_{n+1}-z^s_n$ &9.1167  &9.1416  &9.1668 \\
\hline
$(z^s_{n+1}-z^s_n)/2\pi$ &0.6892  &0.6873 &0.6854 \\
\noalign{\smallskip}\hline\noalign{\smallskip}
\noalign{\smallskip}\hline
\end{tabular}
\caption{Several stable separation distances $z^s_n$ calculated using the full Hamiltonian ($U_H$), the interaction Hamiltonian ($U_{H_I}$), and numerical time-evolution simulations. Parameters as in Fig.~\ref{BS_sketch}. }
\label{tab_inter1}       
\end{table}

\subsection{Wavelength of the oscillatory tails}
Another approach to finding the separation $\Lambda_U$ between consecutive stable separation distances 
of two bound solitons, is to study the {\it spatial dynamics} of the system \cite{Champneys_homoclinic,Gelens_NL}.
Defining the variables $y_1(x)={\rm Re}[A(x)]$,  $y_2(x)={\rm Im}[A(x)]$,  $y_3(x)=d_x {\rm Re}[A]$  and $y_4(x)=d_x {\rm Im}[A]$, (\ref{LLEsteady}) can be recast as the four dimensional dynamical system given by 
\begin{equation}\begin{array}{l}\label{SD_full}
d_xy_1=y_3\\
d_xy_2=y_4\\
d_xy_3=y_2+\theta y_1-y_1y_2^2-y_1^3\\
d_xy_4=-y_1+\theta y_2-y_2y_1^2-y_2^3+\rho.
\end{array}\end{equation}
In this framework, LSs biasymptotic to the HSS $A_0^b$ correspond to {\it homoclinic orbits} to $A^b_0$ \cite{Holmes}. By studying the linearization
of the system (\ref{SD_full}) around $A_0^b$, and therefore the spectrum of eigenvalues of its Jacobian matrix, it is possible to understand how trajectories leave and approach
$A_0^b$. The spatial eigenvalues of the Jacobian matrix are solutions of the characteristic polynomial
\begin{equation}\label{biqua_general}
 \lambda^4+c_2\lambda^2+c_0=0.
\end{equation}
with $c_0=\theta^2+3I_0^2-4\theta I_0+1$ and $c_2=4I_0-2\theta$.
This equation is invariant under $\lambda\rightarrow-\lambda$ and, since the coefficients are real, it leads to eigenvalues whose configuration in the complex plain is symmetric with respect to both axes. The form of this equation is a consequence of spatial reversibility \cite{Devaney_A,Homburg,Knobloch15}. The eigenvalues satisfying (\ref{biqua_general}) are
\begin{equation}
 \lambda=\pm\sqrt{(\theta-2I_0)\pm\sqrt{I_0^2-1}}.
\end{equation}

For any value of $\theta$, $A^b_0$ is a saddle-focus for $I_0<I_c$, with eigenvalues $\lambda_{1,2,3,4}=\pm q_0\pm ik_0$, where
\begin{equation}\label{q0}
q_0=\frac{1}{\sqrt{2}}\sqrt{\sqrt{c_0}+\theta-2I_0}
\end{equation}
and
\begin{equation}\label{k0}
k_0=\frac{1}{\sqrt{2}}\sqrt{\sqrt{c_0}-(\theta-2I_0)}.
\end{equation}
In this case, trajectories approach or leave $A_0^b$ in an oscillatory manner, and therefore, the tails of the soliton profile are oscillatory in space. Thus, in the linear regime, the oscillatory tails of the soliton profile are approximated by 
\begin{equation}\label{F1}
 {\rm Re}[A]={\rm Re}[A_0]+a_1{\rm e}^{q_0 x}{\rm cos}(k_0 x+\varphi_1),
\end{equation}
where $a_1$ and $\varphi_1$ are functions of the control parameters $(\theta,\rho)$ determined by the complete nonlinear dynamics, and they can be estimated by fitting the tails with function (\ref{F1}). For the control parameter values used here, $(\theta,\rho)=(1.5,1.11445)$, the frequency and the decay rate are respectively $k_0=0.6853$ and $q_0=0.1827$, and the fitted parameters are $(a_1,\varphi_1)=(1.45\cdot10^{-7},-0.053)$. The imaginary part of the tails is described by a function analogous to (\ref{F1}).

\begin{figure}[!t]
\centering
\includegraphics[scale = 0.85]{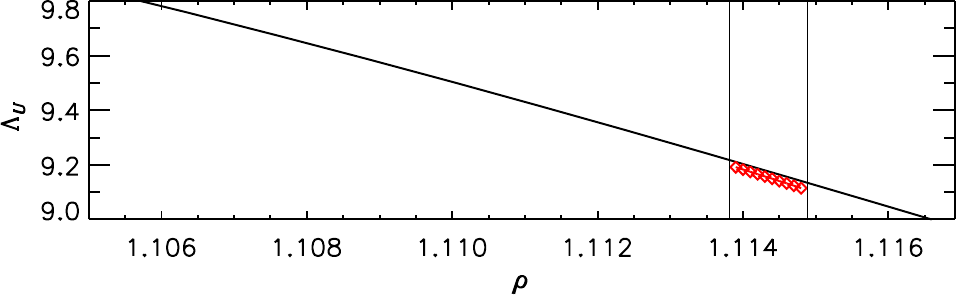}
\caption{The solid line shows the wavelength of the oscillatory tails, calculated using (\ref{analic_wavelength}) for $\theta=1.5$. For comparison, the separation $\Lambda_U$ between stable BSs as calculated using the interaction potential is shown by the red diamonds in the region of existence of LSs.}
\label{periodSOD}
\end{figure}

From (\ref{k0}), one finds that the wavelength of the oscillatory decaying tails is given by 
\begin{equation}\label{analic_wavelength}
\lambda_0^{tails}=\displaystyle\frac{2\sqrt{2}\pi}{\sqrt{\sqrt{c_0}-(\theta-2I_0)}}.
\end{equation}
This expression is plotted in Fig.~\ref{periodSOD} as a function of $\rho$ for $\theta=1.5$. The red diamonds correspond to the periodicity of the potential $\Lambda_{U}$, calculated using (\ref{effec_pot}) in the pinning region of the homoclinic snaking. The fact that $\Lambda_{U}\approx\lambda_0^{tails}$ shows that both approaches (interaction potential vs. spatial eigenvalues) can be used to calculate stable separation distances from an initial stable separation distance $z^s_n$, by using the relation $z^s_{n+1}=z^s_n+\Lambda_{U}$.

\begin{figure}[!t]
\centering
\includegraphics[scale=0.9]{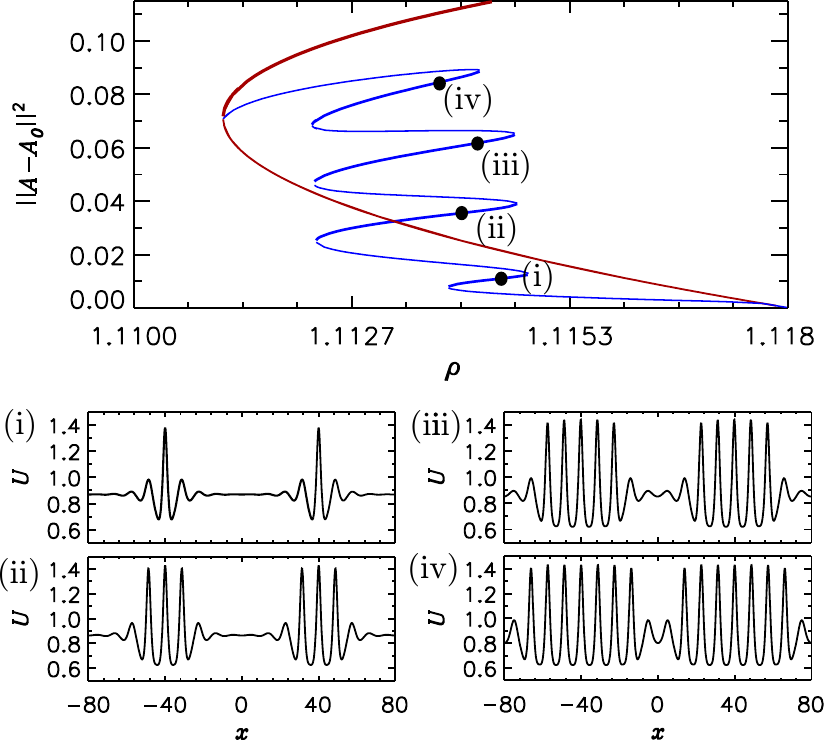}
\caption{(a) Bifurcation diagram of BSs consisting of two equidistant LSs (separated by $L/2$). The profiles indicated on the branches are shown in panels (i)-(iv). We have considered $\theta=1.5$ and $L=160$.}
\label{Sanake2}
\end{figure}

\begin{figure*}[]
\centering
\includegraphics[scale=0.9]{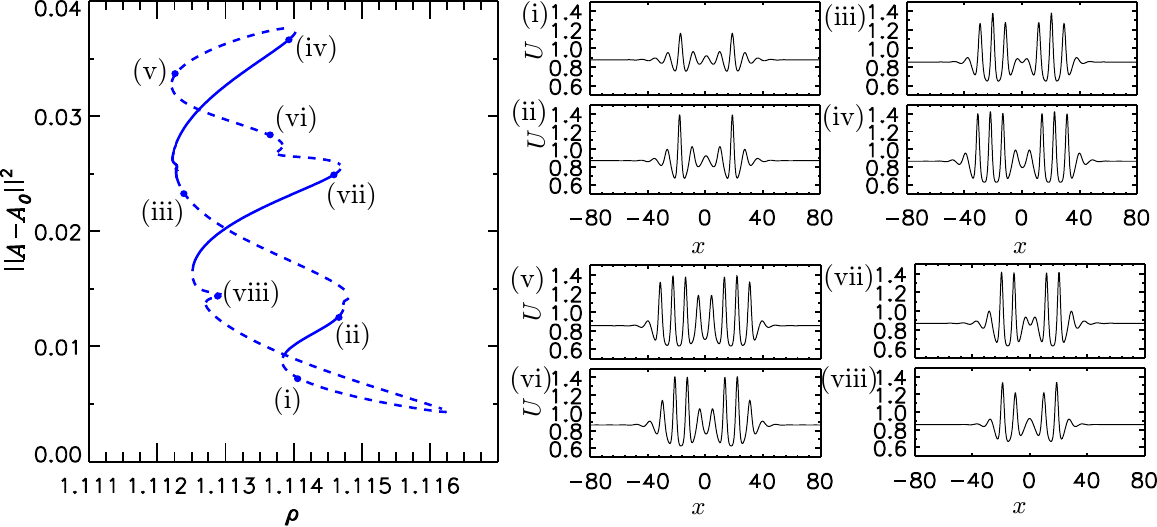}
\caption{An isola formed by BSs consisting of two pulses. Solid (dashed) lines indicate stable (unstable) solutions. Examples of solution profiles
(real part $U = {\rm Re}[A]$) are shown in panels (i)-(viii). Parameters as in Fig.~\ref{Sanake2}.}
\label{isola1}
\end{figure*}

\section{Bifurcation structure}\label{sec:interac:bif}
In this Section, we explore the bifurcation structure of some of the BSs that can be formed combining LSs consisting of one or multiple peaks locked at different separation distances, such as those shown in Fig. \ref{BS_sketch}. A detailed study of the bifurcation structure of BSs has been done in \cite{Multi_Burke} in the context of the Swift-Hohenberg equation. In the previous Section, we showed that 
the equilibrium separations between two soliton states are given by the overlapping of the tail of one soliton with the core of the other one. The presence of stable BSs is intrinsically determined by the properties of the oscillatory tails. Therefore, for a fixed set of parameters $(\theta,\rho)$, one expects to have as many BSs as one can make combinations of existing LSs. In an infinite domain, the number of possible BSs is infinite, while in a finite size system the number of BSs is constrained by the domain size.

First, let us discuss the bifurcation structure of BSs consisting of two single solitons separated by $L/2$. As these pulses are equidistant to the nearest neighbor on either side, their behavior is identical to the behavior of single pulses on a periodic domain of size $L/2$. These type of structures are, therefore, organized in a snaking bifurcation diagram as shown in Fig.~\ref{Sanake2}. Both pulses behave similarly as a single pulse previously shown in Fig.~\ref{Hsnake}, and so each individual pulse of the BS adds extra peaks after each saddle-node bifurcation.

As shown in Fig. \ref{BS_sketch}, BSs not only exist with a stable separation distance $L/2$, but solutions with various other stable separations $z^s_n$ (see Table~\ref{tab_inter1}) exist. Similarly, BSs consisting multiple peaks separated by $z^s_n$ also exist. Such solutions share the same bifurcation structure as the LSs separated 
by $L/2$, shown in Fig.~\ref{Sanake2}, as long as both LSs within the BS remain far enough to only interact weakly and behave as independent entities.

\begin{figure*}[!t]
\centering
\includegraphics[scale=0.9]{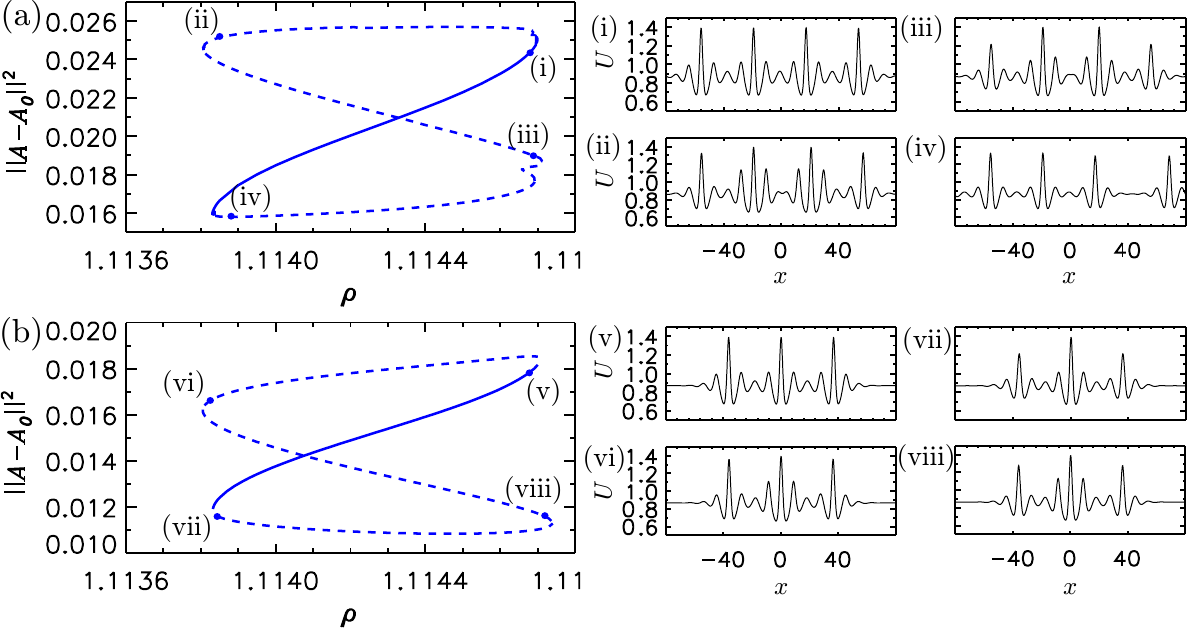}
\caption{Isolas formed by BSs consisting of four (a) and three (b) pulses. Solid (dashed) lines indicate stable (unstable) solutions. 
Examples of solution profiles (real part $U = {\rm Re}[A]$) are shown in panels (i)-(viii). Parameters as in Fig.~\ref{Sanake2}.}
\label{isola2}
\end{figure*}

 The branches associated to BSs corresponding to LSs that are close to one another are, however, organized into large 
isolas, and do not connect directly with the pattern branches. Such a bifurcation structure of isolas is shown 
in Fig.~\ref{isola1} for BSs separated by $z^s_1$. This kind of behavior was reported in \cite{Multi_Burke} 
to originate from insufficient accuracy in the continuation algorithm, which generates
jumps between independent isolas, creating one large one as seen in Fig.~\ref{isola1}. Despite 
improving the accuracy in the numerical continuation algorithm, we have not found such independent isolas in the LL equation, 
instead we always find several connected isolas in a structure similar to that of Fig.~\ref{isola1}. 

In addition to the symmetric BSs consisting of two LSs, there is a wide variety of BSs that are built up from any number of pulses randomly separated by the pinning distances $z^s_n$. Many of these states are not invariant under reflection symmetry, i.e. are asymmetric, and in the LL equation they move with constant velocity as a consequence of non-variational effects.
If the separation between the peaks of these states is large, they form a snaking-type bifurcation diagram (see Fig.~6 in \cite{Multi_Burke}). However, when the separation between LSs is smaller, these BSs are also organized in isolas. Fig.~\ref{isola2} shows two examples of such isolas.  Panel (a) shows the isola corresponding to a four-pulse BS whose peaks are separated by a distance $z^s_2$. Panel (b) shows a similar structure but for a 4-pulse BS where the inter-distance between peaks is also given by $z^s_2$.

\section{Oscillatory tails induced by higher order dispersion}\label{sec:interac:HOT}
In this Section, we study the effect of higher order dispersion on the shape of the oscillatory tails
of solitons, and as a result, on the stability and shape of BSs. Taking into account higher order dispersion effects is critical 
when considering cavities that operate close to zero GVD, a situation which has been targeted experimentally to achieve broader
frequency combs \cite{Lamont,Brasch}. Theoretically, various works have addressed the influence of higher order dispersion
on the formation and stability of solitons \cite{Coen_comb,GelensPRA2007,Tlidi_FOD,Milian1,Pedro_TOD,Milian2}.

\begin{figure}[!t]
\centering
\includegraphics[scale = 0.85]{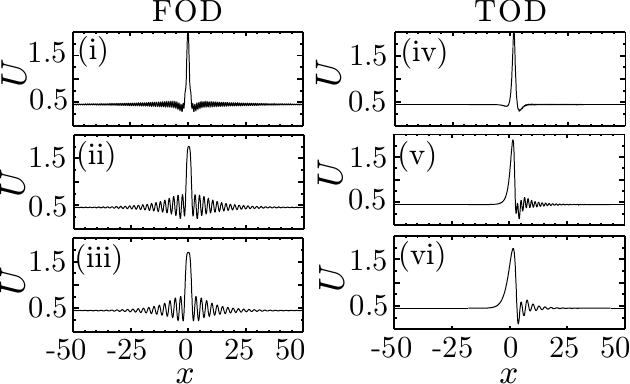}
\caption{Spatial profile of solitons for $\theta=2.2$ and $\rho=1.5$ with different strengths of higher order dispersion. In (i)-(iii) FOD is added with $d_4 = 0.02, 0.1, 0.3$, respectively. In (iv)-(vi) TOD is added with $d_3 = 0.1, 0.3, 0.6$, respectively.}
\label{singleFOD_TOD}
\end{figure}

\subsection{Solitons in the generalized LL equation with higher order dispersion}
The LL equation in the anomalous regime including dispersion terms of third and fourth order reads
\begin{equation}\label{LLEHOD}
 \partial_tA=-(1+i\theta)A+i\partial_x^2A+d_3\partial_x^3A+id_4\partial_x^4A+iA|A|^2+\rho
\end{equation}
Fig. \ref{singleFOD_TOD} shows several stable soliton solution profiles for $(\theta,\rho)=(2.2,1.5)$, including either fourth order dispersion (FOD, $d_4 \neq 0$) 
or third order dispersion (TOD, $d_3 \neq 0$), each time varying the strength of this higher order dispersion term. In both cases, adding these 
dispersion terms of higher order introduces oscillatory tails where previously 
(only second order dispersion, and $\theta>2$) they were highly damped. Initially (small FOD or TOD), 
these oscillations have a small amplitude and very short wavelength, but when increasing the higher
order dispersion strength further these oscillatory tails become more pronounced and their wavelength 
increases. One critical difference between FOD and TOD is the symmetry of the resulting soliton profiles.
While FOD retains reversibility in the system, TOD breaks such reversibility. As a result, 
solitons in systems with FOD remain symmetric, while in the presence of TOD they become asymmetric and start drifting (to the right for $d_3>0$).

\begin{table}
\centering  
\begin{tabular}{llllllllll}
\hline\noalign{\smallskip}
$d_4=0.15$ &  $z^s_4$& $z^s_9$  & $z^s_{16}$& $z^s_{25}$ \\
\hline
 $U_{H_{I}}$&12.7976 &24.2656  &40.3503   &61.0321\\
\hline
Exact&12.6436   &24.2517&40.4122   &61.119 \\

\hline\noalign{\smallskip}
$d_3=0.3$ &  $z^s_0$& $z^s_3$  & $z^s_{7}$& $z^s_{13}$ \\
\hline
 $U_{H_{I}}$&14.788130 &19.357768  &25.453909   &34.600589\\
\hline
Exact&14.966661   &19.517315&25.601455   &34.742282 \\

\noalign{\smallskip}\hline\noalign{\smallskip}
\noalign{\smallskip}\hline
\end{tabular}
\caption{Stable separation distances $z^s_n$ for $\theta=2.2$ and $\rho=1.5$ calculated with the interaction potential are compared with the exact numerical solution, both for a situation with FOD ($d_4=0.15$) and a situation with TOD ($d_3=0.3$).}   
\label{tab2}  
\end{table}

\subsection{Interaction potential}
As before, inserting the ansatz (\ref{ansatz_interac}) into (\ref{effec_pot_int}) the potential $U_{H_I}$ can be calculated as a function of the separation distance $z$. Also in the presence of higher order dispersion, this leads to an oscillatory interaction potential, where the extrema give an estimate of the equilibrium separation distances of solitons within BSs. Such approximation is in principle only valid when the separation between solitons in the BS is large enough. Therefore, in Table \ref{tab2}, we compare the predictions for the separation distances using the interaction potential with the exact numerical solution, finding good correspondence. 

\begin{figure}[!t]
\centering
\includegraphics[scale = 0.9]{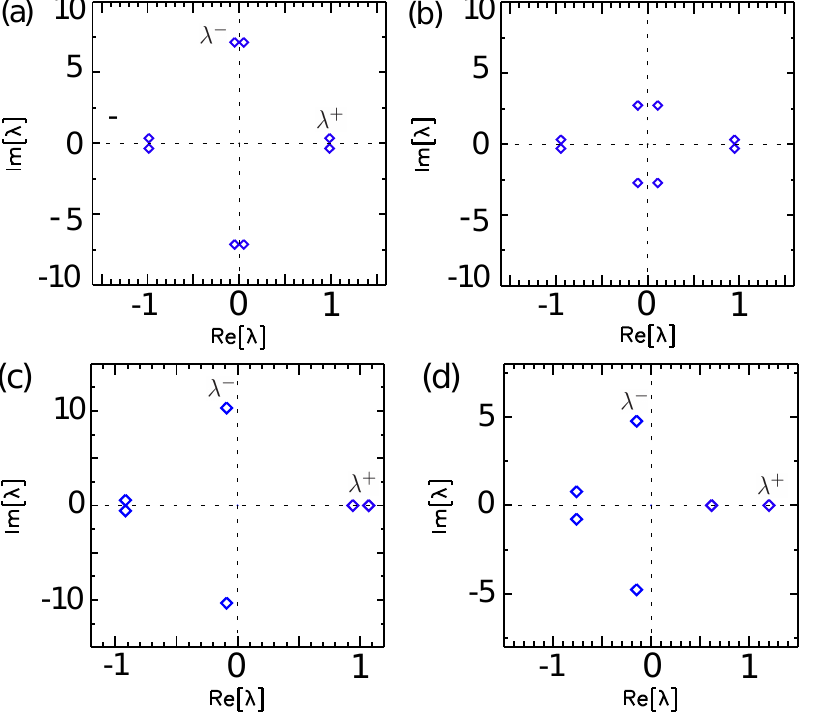}
\caption{Spatial eigenvalues satisfying the characteristic polynomial (\ref{charac_poli_FOD}) for $d_4=0.02$ (a) and $d_4=0.15$ (b), and satisfying the characteristic polynomial (\ref{eigen_TOD}) for $d_3=0.1$ (c) and $d_3=0.25$ (d). We have considered $\theta=2.2$ and $\rho=1.5$.}
\label{eigenvalues_HOD}
\end{figure}

\begin{figure}[!t]
\centering
\includegraphics[scale = 0.9]{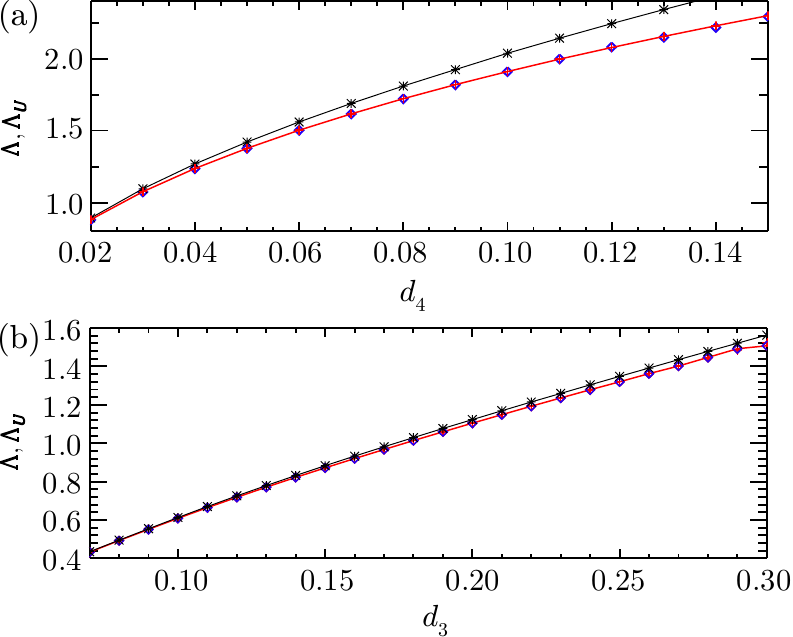}
\caption{Wavelength of the oscillatory tails for $\theta=2.2$ and $\rho=1.5$ (red line with crosses), and, equivalently, the periodicity of the potential $\Lambda_{U}$ (blue diamonds), as function of FOD (a) and TOD (b) coefficients. 
 The results of the analytical approximations (\ref{app_FOD}) and (\ref{app_TOD}) are plotted as a black line with asterisks.}
\label{PeriodHOD}
\end{figure}

\subsection{Wavelength of the oscillatory tails: fourth order dispersion}
Similarly as in Section~\ref{sec:interac:LS}, using the variables $y_1={\rm Re}[A]$, $y_2={\rm Im}[A]$, $y_3=d_x{\rm Re}[A]$, $y_4=d_x{\rm Im}[A]$,
$y_5=d_x^2{\rm Re}[A]$,
$y_6=d_x^2{\rm Im}[A]$, $y_7=d_x^3{\rm Re}[A]$ and $y_8=d_x^3{\rm Im}[A]$, one can recast the stationary version of (\ref{LLEHOD}) to the following dynamical system:
\begin{equation}\label{DS_FOD}
 \begin{array}{l}
  d_xy_i=y_{i+2}, \hspace{0.55cm} i=1,...,6\\
  d_xy_7=d_4^{-1}\left[y_2+\theta y_1-y_5-y_1(y_1^2+y_2^2)\right]\\
  d_xy_8=d_4^{-1}\left[-y_1+\theta y_2-y_6-y_2(y_1^2+y_2^2)\right].
 \end{array}
\end{equation}
The eigenspectrum of the Jacobian of (\ref{DS_FOD}) evaluated at $A_0$ is given by the characteristic polynomial
\begin{equation}\label{charac_poli_FOD}
d_4^2\lambda^8+2d_4\lambda^6+c_4\lambda^4+c_2\lambda^2+c_0=0,
\end{equation}
with $c_4=(1-2d_4\theta+4d_4I_0)$.
The solution of this polynomial consists of two sets of eigenvalues $\lambda_{1,2,3,4}=\pm q^+\pm ik^+$ and $\lambda_{5,6,7,8}=\pm q^-\pm ik^-$, as
those shown in Fig.~\ref{eigenvalues_HOD} for $d_4=0.02$ (a) and $d_4=0.15$(b). Due to the spatial reversibility the eigenspectrum is symmetric with 
respect to the axes Im$[\lambda]=0$ and Re$[\lambda]=0$. This eigenspectrum allows estimating the wavelength of the oscillatory
tails by $2\pi/k^-$, where $k^-$ is the imaginary part of the spatial eigenvalue with smallest negative real part in
absolute value ($q^-$). Fig.~\ref{PeriodHOD}(a) shows that the wavelength of the oscillatory tails increases with $d_4$.
Moreover, it shows that this wavelength (red line with crosses) corresponds well with an estimate of the periodicity of the interaction potential (blue diamonds), again confirming the validity of both approaches. 

For low values of $d_4$, the modulus of the spatial eigenvalue  $\lambda^-$ is very large (see Fig.~\ref{eigenvalues_HOD}), in such a way that the dominant 
terms in (\ref{charac_poli_FOD}) are those with the highest order in $\lambda$. Due to this it is possible to obtain
an analytical approximation for those eigenvalues by solving 
(\ref{charac_poli_FOD}) only considering the highest order in $\lambda$. In this way, we obtain that $\lambda^-$ can be approximated by the expression
\begin{equation}\label{app_FOD}
 \lambda^-=\pm q^-\pm ik^-=\pm\displaystyle\sqrt{\frac{\sqrt{c_4}-1}{2d_4}} \pm i\displaystyle\sqrt{\frac{\sqrt{c_4}+1}{2d_4}}.
\end{equation}
This expression shows that the eigenvalue approaches infinity ($\lambda^-\rightarrow\infty$) when FOD becomes zero ($d_4\rightarrow0$),
a result that can be observed looking at the spectrum for different values of $d_4$. When $d_4\rightarrow0$, these two eigenvalues $\lambda^-$, and its complex conjugate,
tend to $\pm i\infty$. The approximation of the wavelength of the oscillatory tails using this expression (\ref{app_FOD}) gives:
\begin{equation}
	\Lambda=2\pi/k^-=2\pi\displaystyle\sqrt{\frac{2d_4}{\sqrt{c_4}+1}},
\end{equation}
and is shown in Fig.\ \ref{PeriodHOD}(a) by the black line with asterisks.

\subsection{Wavelength of the oscillatory tails: third order dispersion}
Following the analysis for FOD, we now write down the LL equation with TOD and look for solutions that move rigidly at a velocity $v$ (to be determined), namely $A(x,t)=A(x-vt)$. From (\ref{LLEHOD}) with $d_4=0$ one has:
\begin{equation}\label{LLE_moving_solution}
 -v d_xA=-(1+i\theta)A+i d_x^2A+d_3 d_x^3A+iA|A|^2+\rho,
\end{equation}
which as in the previous section can be recast into: 
\begin{equation}\begin{array}{l}\label{SD_TOD}
d_xy_i=y_{i+2},\hspace{0.2in} i=1,...,4\\
d_xy_5=d_3^{-1}[y_4-vy_3+y_1-\theta y_2-y_2y_1^2-y_2^3-\rho]\\
d_xy_6=d_3^{-1}[-y_3-vy_4+y_2-\theta y_2-y_1^3-y_1y_2^2 ]. 
\end{array}\end{equation}
 Steadily drifting soliton solutions of (30) can be
found numerically, with arbitrary precision, using a
Newton-Raphson method where the velocity $v$ is computed as part of the solution.
Then, the eigenspectrum of the Jacobian of (\ref{SD_TOD}) around $A_0^b$ with the found velocity is given by the characteristic equation:
\begin{equation}\label{eigen_TOD}
d_3^2\lambda^6+(2vd_3+1)\lambda^4-2d_3\lambda^3+(c_2+v^2)\lambda^2-2v\lambda+c_0=0.
\end{equation}
The corresponding spatial eigenvalues are shown in Fig.~\ref{eigenvalues_HOD} for $d_3=0.1$ (c) and $d_3=0.25$ (d). As reversibility is broken by TOD, the spatial eigenvalues no longer appear in quadruplets, namely complex eigenvalues appear in conjugate pairs while real eigenvalues appear non-symmetrically around the imaginary axis. 
Similar as in the FOD case, however, the wavelength of the oscillatory tails as determined 
by the dominant spatial eigenvalue $\lambda^-$ increases with TOD strength $d_3$, as shown in Fig.~\ref{PeriodHOD}(b).  

We can obtain an analytical approximation for the wavelength of the oscillatory tails $\Lambda$ using a similar analysis as in the previous 
FOD case, considering the highest order in $\lambda$:
\begin{equation}\label{app_TOD}
	\Lambda=2\pi/k^-=2\pi\displaystyle\frac{d_3}{\sqrt{1+2vd_3}},
\end{equation}
which is again plotted in Fig.\ \ref{PeriodHOD}(b) as a black line with asterisks. We note that the approximation (\ref{app_TOD}) gives very accurate results over a broad range of values of the parameter $d_3$, being quite more accurate than the equivalent approximation in the case of FOD.

\begin{figure*}[]
\centering
\includegraphics[scale =1]{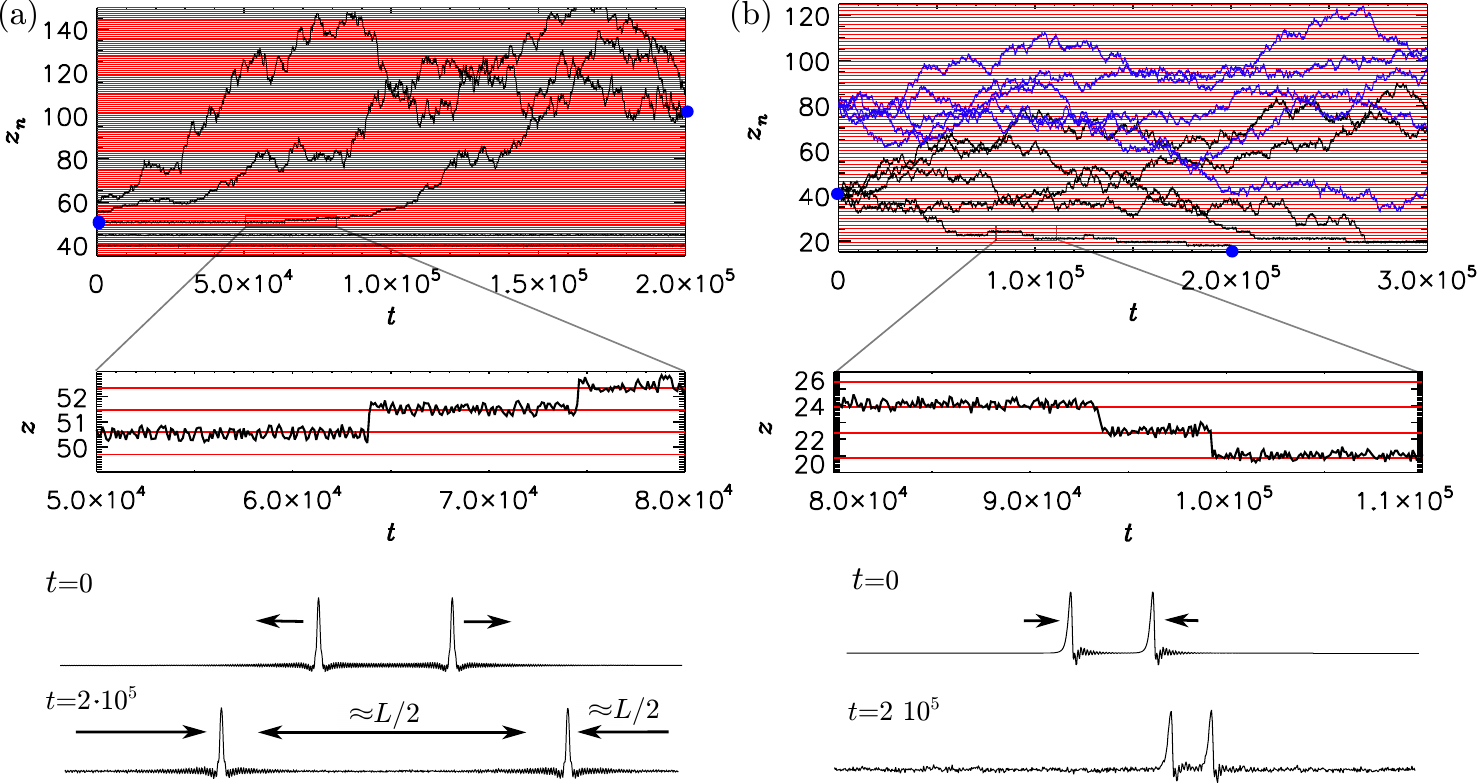}
\caption{Effect of white noise $\xi(x)$ on a BS consisting of two solitons in a reversible system 
(with FOD: $d_4=0.025$ and $\sqrt{D}=0.5$. (a)) and a system with broken reversibility  
(with TOD: $d_3=0.3$ and $\sqrt{D}=1.3$. (b)). The separation distance $z$ between the solitons changes with time,
jumping between stable locking distances ($z^s_n$) indicated by the red lines (and shown in more detail in the zooms).
Each black (blue) line corresponds to a simulation using a given initial separation distance and a different noise realization. 
The initial and final profiles are plotted below for the time points in one trajectory as indicated by blue dots.
We have considered $\theta=2.2$, $\rho=1.5$, and $L=230$.}
\label{noiseFOD}
\end{figure*} 

\section{Effects of noise on bound states}\label{sec:interac:noise}
In this section we study the dynamics of BSs in the presence of noise. The effect of a fluctuating driving term on the dynamics of BSs was studied 
in the NLS equation with a linear dissipative term, where the noise was uniform in space and varying in time \cite{Malomed95random}.
In contrast, in our case, noise is incorporated in the system through
a term that describes a fluctuating pump intensity which is random both in time and space, and that has the form 
\begin{equation}
 \rho=\rho_0+\sqrt{D}\xi(x,t),
\end{equation}
where $\xi(x,t)$ is a Gaussian white noise with zero mean $\langle\xi(x,t)\rangle$=0 and correlations 
\begin{equation}
\langle\xi(x,t)\xi(x',t')\rangle=\delta(t-t')\delta(x-x'),
\end{equation}
where $\langle\cdots\rangle$ stands for the mean value.

In Ref.~\cite{Malomed95random} a Fokker-Planck equation was derived for the probability distribution of the amplitude of a single soliton. Using the stationary solution of such equation, a mean potential for the soliton-soliton interaction was calculated, showing that the bound distances were modified with respect to the deterministic case. 
In our case, the equilibrium separations are not substantially modified and we focus on the jumps between neighboring stable fixed points due to the noise.
\begin{figure*}[t]
\centering
\includegraphics[scale=1]{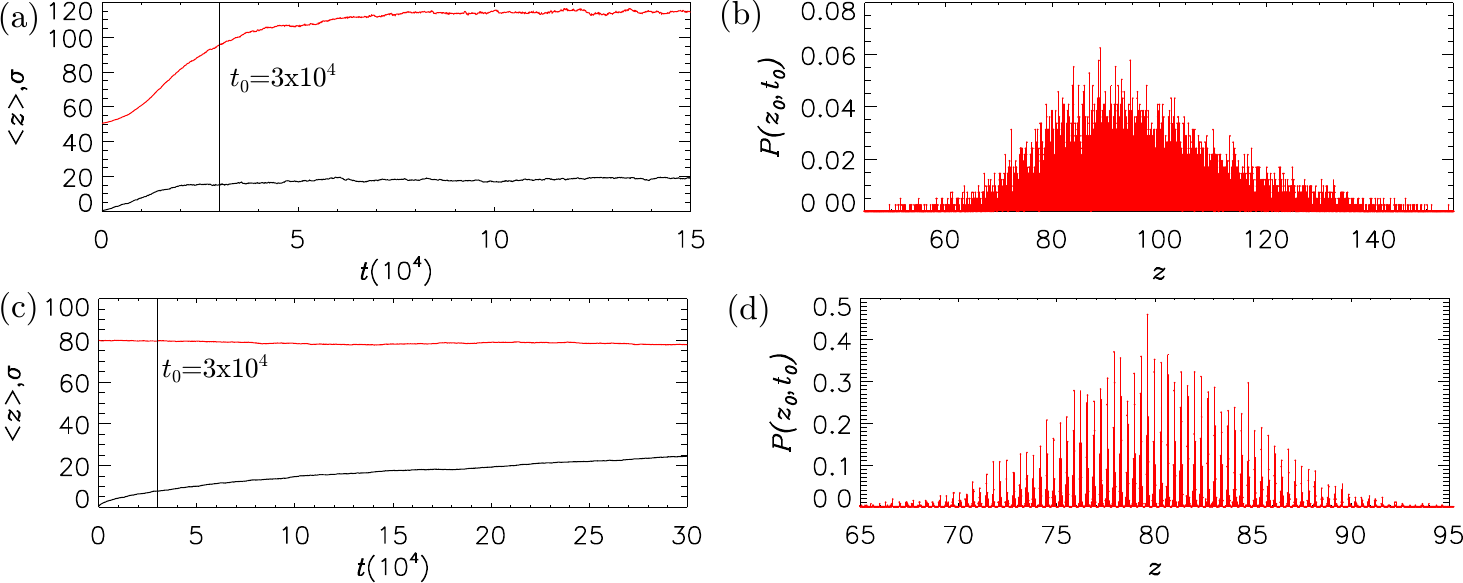}
\caption{On the left, mean $<z>$ (red) and standard deviation $\sigma$ for the separation between peaks $z$ in the case of (a) FOD starting from
$z_0=50$ and (c) TOD starting from $z_0=80$ considering 200 different realizations of the noise. On the right, the probability of finding the
BS with a separation $z$ at time $t_0=3 \times 10^4$ calculated from 13850 different noise realizations for the FOD case (panel (b)), 
and from 8970 for the TOD case (panel (d)).
We use that same parameters as in Fig.~\ref{noiseFOD}.}
\label{mean}
\end{figure*}
Fig.~\ref{noiseFOD} shows the time evolution of BSs consisting of two solitons starting from different initial separations, both in a 
reversible system (with FOD: $d_4=0.025$ (a)) and a system with broken reversibility  (with TOD: $d_3=0.3$ (b)). Due to noise, the separation fluctuates around stable locking distances ($z^s_n$) indicated by the red lines, which is especially clear in the zooms. Sudden larger jumps increasing or decreasing the separation between peaks occur. They correspond to jumps between neighboring equilibrium positions due to the noise. The distance between the stable (red dashed lines)
and unstable (black dashed lines) separations is given by $\Lambda_{U}/2=z^s_{n}-z^u_{n}$, which, for the parameters considered in the figure, corresponds to $\approx0.440$ for the FOD case (a) and  to $\approx 0.754$ for the TOD case (b).
A general trend is that these jumps occur much more often as the separation distance increases, which is expected as the amplitude of the oscillatory tails, and thus the interaction strength, strongly decreases, making the ``potential'' barriers lower. There is, however, a significant difference between the reversible (FOD) system and the system with broken reversibility (TOD). In the reversible system two solitons always tend to increase their separation over long time scales, eventually approaching a BS where both solitons are equidistantly spaced over the domain of size $L$, which is illustrated by the two-soliton profiles taken at the trajectory and times indicated by the blue dots. In the presence of TOD, however, trajectories where two solitons of a BS approach are also commonly observed, and no net drift towards larger or smaller distances is clearly appreciated.


In the case of FOD, Fig.~\ref{noiseFOD}(a) shows two different time scales, a fast one corresponding to the jumps between neighboring stable positions, and a slow one describing how for long times the system evolves to larger separation distances. In an infinite system the separation between the two solitons will increase indefinitely. However, our system is periodic and the largest separation allowed is $z=L/2$. The fact that solitons are not point objects but have a spatial extension allow them to feel the envelope of other solitons. Since in this case solitons move down gradients, two solitons tend to move apart as a result of the interaction through their envelopes. In fact $z=L/2$ correspond to the minimum of such envelope and a minimum of the interaction potential.

In contrast, in the case of TOD solitons move up gradients and it has been shown that the maxima of the interaction potential (and no longer the minima) determine the stable separation distances \cite{Malomed93}. As a result, in principle, one could expect the system now to evolve to smaller separation distances corresponding to the maximum of the envelope of the potential. However, we do not clearly observe such dynamics (see Fig.~\ref{noiseFOD}(b)). Instead no net drift of the separation is observed in our stochastic simulations with TOD. 

The reason behind the different behavior for FOD and TOD can be understood as follows: in addition to the effect of the envelope, another mechanism must be taken into account to explain the dynamics of two interacting solitons in the presence of noise. Note that the wells in the modulated potential shown in Fig.~\ref{Poten} are asymmetric: the barrier to larger distances is lower than to smaller distances. As a result, the noise is rectified and jumps to larger distances happen more often than to shorter distances, inducing a net movement separating the two solitons. This phenomenon is similar to the Brownian ratchet \cite{Magnasco}. This mechanism reinforces the tendency for two solitons to separate in the reversible case (FOD). However, it opposes the envelope effect that pushes solitons towards each other in the presence of TOD (broken reversibility). We observe that the two opposite forces effectively cancel reducing much the net drift, and leading to solitons that seem to wander around randomly.

Considering a large number of different realizations of the noise, and starting from a given initial separation $z_0$, we have calculated the probability $P$ of finding the two solitons
separated a distance $z$ after a time $t_f=30000$ (see Fig.~\ref{mean}). In the case of FOD (panels (a) and (b)) we observe a net drift of the mean position until half of the system size is reached. The variance of the fluctuations grows with the distance due to the potential wells becoming more shallow.
In the case of TOD, the envelope and ratchet effects cancel each other, leading to a nearly diffusive motion of the solitons with zero mean displacement, and variance that grows as the square root of time. The fine structure of the probability distribution also reveals the stable and unstable positions. These are more clear in the case of TOD (Fig.~\ref{mean}d), where for the parameters of the figure the oscillatory tails are more pronounced (see Fig.~\ref{singleFOD_TOD}).

\section{Conclusions}\label{sec:interac:conclusion}

In this work, we have studied the interaction and formation of BSs of localized structures in the LL equation with and without additional higher order dispersion terms. Using variational techniques previously used in  \cite{Malomed91,Malomed93,Malomed94,Akhmediev,Barashenkov}, we have derived an effective potential depending on the separation distance between solitons in the BS. The extrema of this potential determine the stable and unstable separation distances, and they are related with the overlapping of the oscillatory tails of one soliton and the other soliton's core. Therefore, we have also analyzed BSs by studying the eigenvalues of the spatial dynamics for the LL equation, i.e. the way the system approaches or leaves the homogeneous steady state solutions.  The periodicity of the potential is determined by the wavelength of a soliton's oscillatory tails. We have found that the variational approach and the spatial eigenvalue analysis are consistent, and both give accurate estimates of the separation 
between two consecutive equilibrium distances of BSs. Moreover, the variational approach predicts the absolute allowed separation distances.

Next, we have calculated the bifurcation diagrams associated to different types of BSs. BSs that consist of two solitons, which are separated enough such that they interact very weakly through their oscillatory tails, behave as independent LSs, resulting in a snaking-type bifurcation diagram \cite{Multi_Burke}. In contrast, when the separation is smaller such that tail interaction is more significant, BSs are organized in stacks of isolas that are no longer connected to the pattern solution branch. If we consider arrays of more than two solitons, the bifurcation structures obtained are usually isolas.

We have then extended our analysis to the LL equation with higher order dispersion effects, which cannot be ignored when operating close to zero second-order GVD. We considered two qualitatively different situations: (i) fourth order dispersion, which maintains reversibility, and (ii) third order dispersion, which breaks reversibility leading to asymmetric, moving solitons. Both types of higher order GVD can introduce oscillations into the spatial profile of the soliton's tails, even for high values of the detuning, where in the standard LL equation no oscillatory tails were present ($\theta > 2$). By calculating the interaction potentials and characterizing the spatial eigenvalues of the system, we have shown that a wide range of BSs locked at different separation distances come into existence. 

Finally, in Section~\ref{sec:interac:noise}, we have studied the effects of adding white Gaussian noise to the pump intensity, which lead to 
random jumps between BSs of different separation distances. By pooling many time evolution simulations for different noise realizations,
we have calculated the probability of finding two solitons at a given distance after a given time. We have found that there exists a 
critical difference between systems that are reversible (fourth order dispersion) and those that are not (third order dispersion). 
In the reversible case, noise tends to drive the solitons within a BS apart over time, eventually leading to a situation where a BS
consisting of two solitons separated by half of the system width is the most probable one. In contrast, in the non-reversible case, 
the potential envelope and ratchet effect cancel out leading to a diffusive like behavior that leads the two solitons wandering around randomly.

Recently, BSs of solitons have been observed experimentally in semiconductor lasers \cite{Barland}, and in microresonators in the context of frequency comb generation  \cite{Herr_kippenberg,Wang_leo_jae_coen,Wang_leo_jae_coen_Optica,Brasch,Vahala,DelHaye}. Moreover, several binding mechanisms have been studied theoretically and experimentally in the context of passively driven nonlinear optical resonators, such as Gordon/Kelly sidebands, birefringence, and dispersive waves \cite{Wang_leo_jae_coen,Wang_leo_jae_coen_Optica}. We expect that our results will prove useful to help interpreting various of these experimental observations of BSs.

\end{document}